\documentclass[aps,pra,twocolumn,superscriptaddress,groupedaddress,amsmath,amssymb,longbibliography]{revtex4-1}
\usepackage{graphicx}
\usepackage{subfigure}
\usepackage{color}
\usepackage{dcolumn}
\newcommand{\be}{\begin{equation}}
\newcommand{\ee}{\end{equation}}
\newcommand{\bea}{\begin{eqnarray}}
\newcommand{\eea}{\end{eqnarray}}
\newcommand{\ba}{\begin{array}}
\newcommand{\ea}{\end{array}}

\newcommand{\bla}{\color{black}}

\begin{document}

\title{Characterization of quantum dynamics using quantum error
  correction}  \author{S.  Omkar}  \affiliation{Poornaprajna Institute
  of  Scientific Research,  Sadashivnagar, Bengaluru-  560080, India.}
\author{R.            Srikanth}          \email{srik@poornaprajna.org}
\affiliation{Poornaprajna    Institute    of   Scientific    Research,
  Sadashivnagar,   Bengaluru-   560080,  India.}    \affiliation{Raman
  Research   Institute,  Sadashivnagar,   Bengaluru-   560060,  India}
\author{Subhashish   Banerjee}  \affiliation{   Indian   Institute  of
  Technology Jodhpur, Rajasthan - 342011, India}

\begin{abstract}
Characterizing  noisy  quantum  processes   is  important  to  quantum
computation  and  communication  (QCC),   since  quantum  systems  are
generally open.  To  date, all methods of  characterization of quantum
dynamics (CQD),  typically implemented by quantum  process tomography,
are \textit{off-line},  i.e., QCC and  CQD are not concurrent,  as they
require  distinct state  preparations.   Here we  introduce a  method,
``quantum error  correction based  characterization of  dynamics'', in
which  the initial  state is  any  element from  the code  space of  a
quantum  error  correcting  code  that  can  protect  the  state  from
arbitrary  errors acting  on the  subsystem subjected  to the  unknown
dynamics.  The  statistics of  stabilizer measurements,  with possible
unitary pre-processing operations, are used to characterize the noise,
while the  observed syndrome can be  used to correct the  noisy state.
Our method requires at  most $2(4^n-1)$ configurations to characterize
arbitrary noise acting on $n$ qubits.
\end{abstract}
\maketitle

\section{Introduction}
The   principal  difficulty   in   implementing  quantum   computation
physically is  environment-induced noise, which  decoheres the quantum
system, resulting  in the loss  of superposition and  of entanglement.
The noise acting  on a quantum system starting  initially in a product
state with its environment is  described by a completely positive (CP)
map and  is represented by the Kraus  operators \cite{NC00} $E_j
\equiv  \sum_i \alpha_{i,j}F_i$,  where $F_i$  is an  element  from an
operator  (or  error)  basis  satisfying the  orthogonality  condition
Tr$(F_iF^\dag_j)=d\delta_{i,j}$, where $\delta_{i,j}$ is the Kronecker
delta, and $d=2^p$  is the dimension of the  system, consisting of $p$
qubits.  Thus, if $\rho$ represents the initial quantum state, then
\begin{eqnarray}
\mathcal{E}(\rho)      &=&     \sum_j      E_j\rho      E^\dag_j     =
\sum_{m,n} \chi_{m,n}F_m \rho F_n^\dag
\label{eq:chi}
\end{eqnarray}
where  $\chi_{m,n} \equiv  \sum_j \alpha_{j,m}\alpha^\ast_{j,n}$  is a
Hermitian  matrix (the  ``process matrix'')  in  the $d^2$-dimensional
Hilbert-Schmidt  space of  linear operators  acting on  the  system of
dimension   $d$.    From   the   completeness   condition,   we   have
$\sum_{j=1}^{d^2} E_j^\dag  E_j = \sum_{m,n}  \chi_{m,n} F^\dag_mF_n =
\mathbb{I}$, which imposes $d^2$ conditions, so that the matrix $\chi$
has $d^4-d^2$  independent real elements.  Since taking  trace on both
sides yields  $\sum_j \chi_{j,j}=1$, the  (positive) diagonal elements
of $\chi$  can be interpreted  as probabilities.  In this  work, $F_j$
are multi-qubit  Pauli operators,  which is appropriate  for employing
the QEC formalism.

The  characterization of noisy  quantum processes,  namely determining
the matrix  elements $\chi_{m,n}$, was initially  addresed by standard
quantum process tomography  (SQPT) \cite{NC00,DAr00}.  Here the system
undergoing  the  unknown  noisy  dynamics  is  initially  prepared  in
suitable states  and subjected  to state tomography  measurements.  In
ancilla-assisted   process   tomography   (AAPT)  \cite{ABJ+03},   the
principal system  $\textbf{P}$ and an ancillary  system \textbf{A} are
prepared  in  suitable  initial  states,  and  information  about  the
dynamics of $\textbf{P}$ is  extracted via quantum state tomography on
the joint system using  separable or non-separable basis measurements.
SQPT  and AAPT  are  indirect in  that  they first  obtain full  state
tomographic data $\mu_{m,n} = \textrm{Tr}(\rho_m \mathcal{E}(\rho_n))$
on  input states $\rho_n$,  and then  invert this  exponentially large
data  (of   size  $d^4-1$  in   SQPT  and  $d^4-d^2$  in   AAPT  under
trace-preserving noise) to derive $\chi$.

By  contrast,  direct  characterization  of  quantum  dynamics  (DCQD)
\cite{ML06,ML07},  bypasses the  state  tomography.   It uses  quantum
error detection measurements augmented by normalizer measurements in a
code-space  determined  by  stabilizers  corresponding  to  Bell-state
measurements.  Other recent  developments include the characterization
of noise using  an efficient method for transforming a  channel into a
symmetrized  (i.e.,  having  only  diagonal elements  in  the  process
matrix) channel  via twirling \cite{ESM+07}, suitable  for identifying
quantum error correcting codes (QECCs) \cite{SMK+08}.  Recently, three
independent  proposals have  been  presented to  rapidly estimate  the
channel using quantum error  correction (QEC) techniques \cite{FSK+14,
  CFC+14,  Fuj14}, which  aim for  concurrent preservation  of quantum
information, rather  than for  process tomography  of the  dynamics of
\textbf{P}.   A  method  similar  to \cite{ESM+07},  but  extended  to
efficiently estimate  any given  off-diagonal term, was  introduced in
Ref. \cite{BPP08}.

Suppose  we  have  a  situation  where it  is  known  with  reasonable
confidence that an  arbitrary noise is restricted to  a certain known,
sufficiently   small   subsystem   of  a   quantum   computation   and
communication (QCC) device, say a quantum computer.  One can construct
QECCs that  would protect against the  noise.  On the other  hand, the
statistics  of  the  measured  syndrome outcomes  could  be  used  for
characterization  of  that noise,  which  could  be useful  for  other
quantum information processing tasks.  A method that helps combine QCC
and characterization of quantum dynamics  (CQD) would thus surely help
save  valuable quantum  resources.  In  this work,  we present  such a
method, a QEC-based characterization of quantum dynamics (QECCD).  The
reason that the  noise must be restricted to a  known subsystem of the
quantum computer is  that the allowed errors must form  a group, for a
reason  that   will  become   clear  later.   Without   the  subsystem
restriction, our  method can still  be used to determine  the diagonal
terms of the process matrix in the Pauli representation.

From the perspective of CQD, our method allows initial states that are
not fixed but, instead, can be any  thing in the code space of a QECC.
This means that  the noise characterization is  indifferent to certain
kinds of errors in state preparation, namely those that keep the state
within  the  stabilized  code  space\bla.   Our  method  is  presently
restricted to CP--but not necessarily trace-preserving--maps, though
the QEC formalism is known \cite{SL09} to be applicable even to non-CP
maps.

The    remainder   of    this   work    is   as    follows.    Section
\ref{sec:QECC2noise} presents the basic  motivation for using QECC for
CQD.   The  basic  intuition  here  is  an  isomorphism  that  can  be
established between the allowed noise and the erroneous version of the
logical state.   In Section \ref{sec:stabi}, we  introduce a different
type of stabilizer  codes that are suitable for CQD.   These are QECCs
that correct all  possible errors that occur on  known coordinates and
form  a group.   Here we  give an  example of  a five-qubit  QECC that
corrects  all errors  on  the  first two  qubits,  and furthermore  is
perfect  (i.e., it  saturates  the quantum  Hamming  bound).  In  Sec.
\ref{sec:diag}, we show how the statistics of syndrome outcome data on
this kind of QECCs  can be used to read off the  diagonal terms of the
process matrix.  Accessing off-diagonal terms  is a bit more involved.
In principle,  a suitable unitary  can be used to  rotate off-diagonal
terms in such  a way that a syndrome measurement  can access them.  We
show how this is done in Sec. \ref{sec:offdiag}.  However, this method
can only access the real or  imaginary part of off-diagonal terms.  In
Sec.  \ref{sec:toggle}, we  show how a ``toggling''  can be customized
to the  above unitary, such that  the real and imaginary  parts of the
accessed off-diagonal terms can be `toggled', i.e., exchanged, so that
after toggling, the method of Sec.  \ref{sec:offdiag} can be used.  In
Sec. \ref{sec:exp},  we consider  experimental aspects.  We  point out
that  various  QECCD experiments  are  well  within  the reach  of  an
experimental facility (NMR, quantum  optics, etc.)  where entanglement
generation  and manipulation  are  done.   An example  of  QECCD of  a
single-qubit   noise,  that   would  be   suitable  for   experimental
implementation, is worked out in detail.   To this end, we introduce a
new  three-qubit  perfect stabilizer  code,  which  is applied  to  an
amplitude damping channel on the first qubit.  Finally, we conclude in
Sec.  \ref{sec:conclu}.

\section{Noise characterization and QECCs \label{sec:QECC2noise}}

Like DCQD, our method is direct and requires initial entangled states.
However,  unlike  DCQD  and  other quantum  process  topography  (QPT)
methods,  QECCD requires  no  special initial  state preparation:  any
state in  the $2^k$-dimensional code  space of an  $[[n,k]]$ $n$-qubit
stabilizer code  for QEC  is appropriate,  provided the  code corrects
arbitrary errors on  $m$ $(< n)$ known coordinates  of \textbf{P}. The
syndrome  obtained from  the  stabilizer measurement  can  be used  to
correct  the  noisy state,  while  the  experimental probabilities  of
syndromes will contain information about the noise channel.

We  recollect  that  the  code  is  a  subspace  $\mathcal{C}$,  whose
projector $\Pi_\mathcal{C}$  satisfies the error  correcting condition
$\Pi_\mathcal{C}       F^\dag_a       F_b      \Pi_\mathcal{C}       =
C_{ab}\Pi_\mathcal{C}$,   where  $C_{ab}$   is   a  Hermitian   matrix
\cite{Got09}.  In the case  of non-degenerate QECCs (where $C_{ab}$ is
non-singular), this  defines a  bijective mapping between  the allowed
noise  channel  and  states  in  the  \textit{error  ball}  about  any
QECC-encoded  state  $|\Psi_L\rangle$,  akin  to  a  Choi-Jamiolkowski
isomorphism   \cite{OSB0}.    This   follows   from   the   one-to-one
correspondences:
\begin{equation}
\mathcal{E}   \longleftrightarrow  \{\chi_{m,n}\}  \longleftrightarrow
\sum_{m,n} \chi_{m,n}|\Psi_L^m\rangle\langle\Psi_L^n|,
\label{eq:CJ}
\end{equation}
where the first  correspondence follows by definition,  and the second
from  the fact  that  $\{|\Psi^m_L\rangle \equiv  F_m|\Psi_L\rangle\}$
forms a basis in the error  ball about $|\Psi_L\rangle$.  QECCD can be
seen as exploiting the QECC  isomorphism to determine matrix $\chi$ in
that               various               measurements               on
$\mathcal{E}(|\Psi_L\rangle\langle\Psi_L|)$, the noisy  version of the
initial logical  state $|\Psi_L\rangle$,  will suffice to  extract all
information about $\mathcal{E}$, while extracting no information about
the encoded  state $|\Psi\rangle$.   This result  is non-trivial,
since such an isomorphism exists  quite generally for arbitrary QECCs,
but  the  experimental  accessibility  of off-diagonal  terms  of  the
process  matrix  in  the  Pauli representation  is  possible  in  this
approach only  when the allowed  errors form  a group. Thus  a general
QECC cannot necessarily serve QECCD.

\begin{figure}
\includegraphics[width=0.45\textwidth]{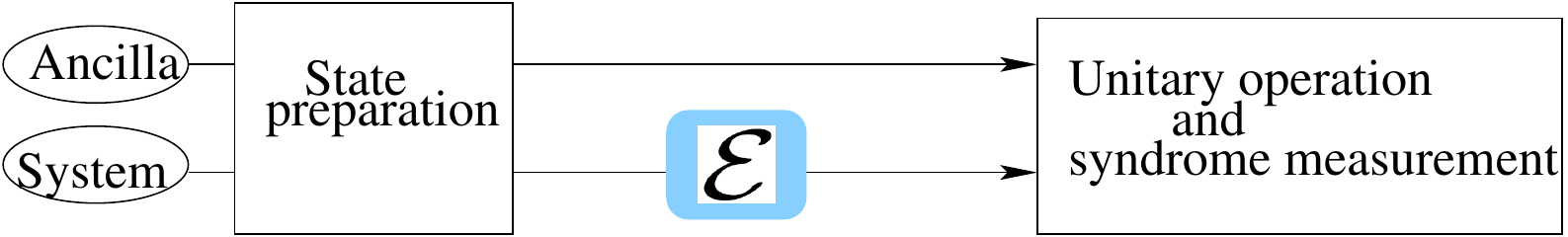}
\caption{(Color  online)  Scheme  for   QECCD:  The  principal  system
  \textbf{P}  (subjected to  the uncharacterized  noise $\mathcal{E}$)
  plus CQD ancilla  \textbf{A} (assumed to be  noiseless) are prepared
  in a QECC-encoded  state.  After \textbf{P} is  subjected to channel
  $\mathcal{E}$   (assumed  correctable   by  the   QECC  used),   the
  stabilizers  are measured  on the  joint system,  possibly following
  unitary operation $U$ or $US^+$.}
\label{fig:qecpt}
\end{figure}

The scheme  for QECCD is  depicted in Fig.  \ref{fig:qecpt}.   Some of
the qubits will  be allowed to be  noisy and others are  assumed to be
clean.  The former qubits constistute the principal system \textbf{P};
the latter the CQD ancilla \textbf{A}.

Suppose  the  full  system  $\textbf{P}+\textbf{A}$ is  in  the  state
$|\Psi_L\rangle  \equiv \sum_{j=0}^{2^k-1}  \beta_j|j_L\rangle$, where
$\{|j_L\rangle\}$  denotes a  logical basis  for the  code space  of a
$[[p+q,k]]$ QECC (which encodes $k$ qubits into $n \equiv p+q$ qubits)
such that  allowed errors in  the $p$ known coordinates  of \textbf{P}
can  be  detected  and  corrected.   An assumption  here  is  that  no
(appreciable)  errors   occur  on  the  $q$   ancillary  qubits.   The
stabilizers  $S_j$ are  a  set of  $p+q-k$  mutually commuting  binary
$n$-qubit   observables   that  stabilize   the   code  space   (i.e.,
$S_j|j_L\rangle  = |j_L\rangle$).  Correctable  errors $F_i$  are such
that for any  pair $F_i, F_j$ ($i\ne j$), there is  at least one $S_i$
that anti-commutes  with the product $F_iF_j$.  This  ensures that the
eigenvalue  pattern for  each correctable  error, which  is  the error
syndrome, is distinct. The Hamming  bound \cite{Got09} in this case is
given by $2^k V \le 2^{p+q}$, where $V$ is the size of the error ball,
here  the set  of all  possible  errors in  \textbf{P}, so  that $V  =
\sum_{r=0}^p  {p  \choose  r}  3^r\cdot1^{p-r}  = 4^p  =  d^2$  (since
$d=2^p$). The Hamming bound for QECCD is thus $q \ge k+p$.

\subsection{A class of stabilizer codes suitable for CQD
\label{sec:stabi}}

To see the connection between QEC and CQD, consider the $[[5,1]]$ code
that saturates the  Hamming bound for an  arbitrary single-qubit error
on any qubit:
\begin{eqnarray}
|0_L\rangle &=& \frac{1}{2\sqrt{2}}  (|00000\rangle  +
|00110\rangle  +  |01001\rangle  -  |01111\rangle 
\nonumber \\ &&~~~-  |10011\rangle  +
|10101\rangle  +   |11010\rangle  +  |11100\rangle)\nonumber\\
|1_L\rangle  &=&
XXXXX|0_L\rangle,
\label{eq:5qecc}
\end{eqnarray}
where the states  are represented in the computational  basis, and $X$
is the Pauli-$X$  operator. We note that the above  code words satisfy
the error correcting conditions when  the allowed errors are arbitrary
errors  on the  first  two qubits.   Thus, let  the  first two  qubits
constitute  \textbf{P},  subjected  to  unknown  dynamics,  while  the
remaining three are CQD ancillas.  There are 16 basis elements for the
general  noise acting  on  these two  qubits,  represented by  $X^{\bf
  u}Z^{\bf v}$, where ${\bf  u}=(u_1,u_2)$ and ${\bf v}=(v_1,v_2)$ are
vectors defined over $GF(2)$.

The stabilizer generators
\begin{equation}
\mathcal{G} \equiv \{IZZZZ, XXXII,  ZXZIX, ZZXXI\}
\label{eq:5G}
\end{equation}
uniquely  determine the  four syndromes  to be  $(u_2, v_1\oplus  v_2,
u_1\oplus  v_2, u_1\oplus  u_2)$.   It is  worth  stressing that  code
(\ref{eq:5qecc}) is different from  that in Ref. \cite{LMP+96} because
the  stabilizers,  and  thus  the   set  of  correctable  errors,  are
different, even  though the code words  are the same.  The  main point
for QECCD  is that the set  $\mathbb{E}$ of correctable errors  (up to
scalar factors  $\pm1$ and $\pm  i$), form a group,  the \textit{error
  group}.  This  is reflected  in the above  Hamming bound  for QECCD.
Suppose  the unknown  dynamics is  a (correlated)  noise given  by the
Kraus   operators   $\{\sqrt{1-p}I,   \sqrt{p}X_1X_2\}$.    From   Eq.
(\ref{eq:chi}) one finds  that the probability that  no error happens,
and   thus  that   to  experimentally   find  the   no-error  syndrome
$(1,1,1,1)$,   is    $\chi_{I,I}=1-p$.    Similarly,    the   syndrome
$({-1},1,{-1},1)$   for  error   $X_1X_2$   occurs  with   probability
$\chi_{X_1X_2,X_1X_2}  = p$.   The syndrome  carries information  only
about the noise, and nothing about  the encoded state, and can be used
to  correct the  noisy version  of $|\Psi_L\rangle$,  while the  error
statistics  determined by  the syndrome  outcomes helps  determine the
elements of matrix $\chi$. (There  are no off-diagonal terms of $\chi$
for this channel in the Pauli operator representation.)

Now  consider a  variant of  the  above example,  wherein we  consider
letting \textbf{P} be all 5 qubits, while  the noise is taken to be an
arbitrary  one-qubit  error  on  any  one qubit.   This  is  just  the
five-qubit  code of  Ref. \cite{LMP+96}.  Though the  above five-qubit
QECC is  suitable for QEC  here, still  the correctable errors  do not
form a closed  set and, thus, do not constitute  an error group: e.g.,
while  $X_1$  and  $Y_2$  can be  corrected,  their  product  $X_1Y_2$
cannot. Although the diagonal terms of the process matrix can still be
calculated,  for  the off-diagonal  terms,  our  method requires  this
closure property.

\subsection{Determining the diagonal terms of $\chi_{m,n}$
\label{sec:diag}}

Given $\mathcal{E}$ known  to be correctable by  a non-degenerate QECC
$Q$, but otherwise uncharacterized, a single configuration suffices to
determine all  diagonal elements  $\chi_{m,m}$ via measurement  of the
(mutually   commuting)  stabilizers   of   $Q$.   We   refer  to   the
corresponding  observable as  the  syndrome  operator, $\Sigma$.   The
measurement of  syndrome $x$, corresponding to  error $F_x$, collapses
the noisy state into the  pure state $F_x|\Psi_L\rangle$, which can be
corrected by  applying $F^\dag_x=F_x$.   The probability  of obtaining
outcome $x$ is:
%\begin{widetext}
\begin{eqnarray} 
&&\xi(x) =
\textrm{Tr}\left(\mathcal{E}\left(
    |\Psi_L\rangle\langle\Psi_L|\right)
   \left[\sum_{J=0}^{2^k-1}|J^x\rangle\langle
  J^x|\right] \right)  \nonumber \\
&&= 
 \langle\Psi_L^x| \left[\sum_{m,n}^{d^2-1}\chi_{m,n}
  |\Psi_L^m\rangle\langle \Psi_L^n|
   \right]
    |\Psi_L^x\rangle  \nonumber \\
&&= \sum_{m,n=0}^{d^2-1} 
\chi_{m,n}
\delta_{x,m}
\delta_{x,n} = \chi_{x,x},
\label{eq:diag}
\end{eqnarray}  
%\end{widetext}
where it is convenient to take  the tracing basis to be any completion
of $\{F_j|\Psi_L\rangle\}$.
 
\subsection{Determining the off-diagonal terms of $\chi_{m,n}$
\label{sec:offdiag}}

Off-diagonal  terms are  obtained  by pre-processing  the noisy  state
using  a  unitary $U$  or  $US^+$,  prior to  stabilizer  measurement.
[Equivalently,  measurements  are  made  in  one  of  two  bases:  the
  ``rotated  basis'' $U\Sigma  U^\dag$  or the  ``toggled and  rotated
  basis'' $(US^+)\Sigma(US^+)^\dag$, as explained below.]  Here again,
the state just after measurement will be $F_x|\Psi_L\rangle$, for some
correctable $F_x$.  Consider a unitary  operator $U(a,b) = \frac{F_a +
  F_b}{\sqrt{2}}$, where  allowed errors $F_a$ and  $F_b$ anti-commute
(else,  we  choose  $U  = \frac{F_a  +  iF_b}{\sqrt{2}})$,  such  that
$F_aF_x$  and   $F_bF_x$  represent   correctable  errors.    This  is
guaranteed by  choosing a QECC  whose correctable Pauli errors  form a
group (up to a scalar factor  $\pm1$ or $\pm i$) under multiplication.
This requirement is met, as in  the first example above, by choosing a
QECC  that corrects  arbitrary  errors on  subsystem \textbf{P}.   Let
$g_AF_A  \equiv F_aF_x$,  where  $F_A$  is a  Pauli  operator and  the
\textit{Pauli  factor}  $g_A  \in  \{\pm1,\pm  i\}$.   Similarly,  let
$g_BF_B \equiv  F_bF_x$.  For  example, if  $F_a = X,  F_x =  Y$, then
$F_A=  Z$ and  $g_A=+i$.  If  $g_A$ and  $g_B$ are  both real  or both
imaginary,  then  we say  that  the  Pauli  factors  are of  the  same
\textit{type}.  If one  of $g_A$ and $g_B$ is imaginary  and the other
real, we say that the Pauli factors are of distinct type.

Operation   $U(a,b)$  rotates   one  correctable   state  to   another
correctable  state.   This alters  the  statistics  of the  stabilizer
measurement without affecting the  correctability.  The probability of
finding the syndrome corresponding to error $F_x$ is now:
\begin{eqnarray} 
&& \xi(a,b,x)\equiv \nonumber \\
&& \textrm{Tr}\left(U(a,b)\mathcal{E}\left(
    |\Psi_L\rangle\langle\Psi_L|\right)(U(a,b))^\dag
   \left[\sum_{J=0}^{2^k-1}F_x|J\rangle
  \langle J|F_x\right] \right) \nonumber \\
 &&= \frac{
 \sum_{m,n}^{d^2-1}\chi_{m,n} \langle F_x
   \left(F_{a \star m} + F_{b\star m}\right)
  \rangle_L\langle 
   \left(F_{n\star a} + F_{n\star b}\right)
    F_x \rangle_L}{2} \nonumber \\
&&=
\frac{1}{2}
 \sum_{m,n}^{d^2-1}\chi_{m,n}
 \left(g^\ast_A\delta_{A,m} + g_B^\ast \delta_{B,m}\right)
  \left(g_A\delta_{A,n} + g_B\delta_{B,n}\right) \nonumber \\
 &&= \frac{\chi_{A,A} + \chi_{B,B}}{2} +
       \frac{g_A^\ast g_B \chi_{A,B} + g_A g_B^\ast\chi_{B,A}}{2}\nonumber\\
&&= \frac{1}{2}\left(\chi_{A,A} + \chi_{B,B}\right) +
   \textrm{Re}\left(g_A^\ast g_B\chi_{A,B}\right),
\label{eq:offdiag}
\end{eqnarray}
where we have used the notation $F_{m \star n} \equiv F_mF_n$, and the
expectation   value  $\langle\cdots\rangle_L$   is  with   respect  to
$|\Psi_L\rangle$.   The first  term  in the  final  expression of  Eq.
(\ref{eq:offdiag}) contains  only diagonal  elements of  $\chi$, which
are determined  by stabilizer measurements without  the application of
any  pre-processing unitaries.   It follows  from the  second term  in
(\ref{eq:offdiag}) that if $g_A$ and $g_B$ are of the same (different)
type, then $\xi(a,b,x)$  depends only on the real  (imaginary) part of
$\chi_{A,B}$.  For  example, suppose  $a=X, b=Y,  x=Z$, in  which case
$g_A  =  -i$  and  $F_A=Y$,  while $g_B=  i$  and  $F_B=X$.   Thus  an
application of $U(X,Y)$ followed by  a $Z$-error syndrome extracts the
real   part   of   $\chi_{X,Y}$.    In   particular,   $\xi(X,Y,Z)   =
\frac{1}{2}(\chi_{X,X} + \chi_{Y,Y})  - \textrm{Re}(\chi_{X,Y})$. Note
that the  state obtained after measurement  in Eq.  (\ref{eq:offdiag})
is  $\rho_f =  |\Psi_L^x\rangle\langle  \Psi_L^x|\left( U  \mathcal{E}
|\Psi_L\rangle\langle\Psi_L|  U^\dag  \right)  |\Psi_L^x\rangle\langle
\Psi_L^x| = \xi(a,b,x)|\Psi_L^x\rangle\langle\Psi_L^x|$,  that is, the
use of  $U$ does not  alter the QEC  procedure, but only  modifies the
error statistics  to be dependent  on off-diagonal terms  according to
the choice of $U$.

If   $F_a$   and  $F_b$   do   not  commute,   then   $U   =  (F_a   +
iF_b)/\sqrt{2}$. In place of Eq. (\ref{eq:offdiag}) we obtain:
\begin{eqnarray}
\xi(a,b,x) &=& \frac{1}{2}(\chi_{A,A} + \chi_{B,B} + i\left[g_A g_B^\ast
  \chi_{B,A} - g_A^\ast g_B\chi_{A,B}\right]) \nonumber \\
 &=& \frac{1}{2}\left(\chi_{A,A} + \chi_{B,B}\right) +
   \textrm{Im}\left(g_A^\ast g_B\chi_{A,B}\right).
\label{eq:offdiag+}
\end{eqnarray}
It follows from  the second term in (\ref{eq:offdiag+})  that if $g_A$
and $g_B$ are of the  same (different) type, then $\xi(a,b,x)$ depends
on the  imaginary (real) part  of $\chi_{A,B}$.  For  example, suppose
$a=I, b=Y,  x=I$, in which case $g_A  = 1$ and $F_A=I$  while $g_B= 1$
and  $F_B=Y$.  An  application of  $U(I,Y)$ followed  by  the no-error
syndrome  is a  function of  the imaginary  part of  $\chi_{I,Y}$.  In
particular,  $\xi(I,Y,I)  =  \frac{1}{2}(\chi_{I,I}  +  \chi_{Y,Y})  +
\textrm{Im}(\chi_{I,Y})$,  where $\chi_{I,I}$  is  the probability  of
obtaining no error.

In  general,  this   will  leave  the  real  or   imaginary  parts  of
off-diagonal terms undetermined. In the  first example above, the only
other measurements  that can  extract information on  $\chi_{X,Y}$ are
the no-error  outcome in  the $U(X,Y)$  configuration [i.e.,  the term
  $\xi(X,Y,I)$] and  the $X$- and  $Y$-error outcomes in  the $U(I,Z)$
configuration [i.e., the terms  $\xi(I,Z,X)$ and $\xi(I,Z,Y)$], all of
which can yield only information about $\textrm{Re}(\chi_{X,Y})$.

\subsection{Toggling operation \label{sec:toggle}}

We solve  this problem by  pre-processing the noisy state  as follows.
Let   $S   \equiv   \textrm{Diag}\left(e^{i\theta_0},   e^{i\theta_1},
e^{i\theta_2},  \cdots, e^{i\theta_{V-1}}\right)$  be a  $V  \times V$
diagonal matrix where $\theta_j \in \{\pm \frac{\pi}{4}\}$, with equal
entries of both signs. Prior to $U$, we apply the operation
\begin{equation}
S^+   =  \bigoplus_{J=0}^{2^k-1}   S_J   \oplus  \mathbb{I}^\prime   =
\sum_{m=0}^{V-1}  \left[   e^{i\theta_m}  \sum_J  |J^m_L\rangle\langle
  J^m_L|\right] \oplus \mathbb{I}^\prime,
\label{eq:S+}
\end{equation}
where $S_J$  is the $S$  gate acting on  the error space of  the $J$th
code  word and $\mathbb{I}^\prime$  is the  identity operation  on the
space $\Xi$ of states lying outside  the error ball of all code words.
From the perspective of experiment
\begin{equation}
S^+         =        \exp\left(i\left\{        \bigoplus_{J=0}^{2^k-1}
\left[\epsilon\left(\sum_{\{m,n\}=0}^{V/2}\sigma^z_{(J^m,J^n)}
  \right)\right]_J \oplus 0\cdot\mathbb{I}^{\prime}\right\}\right),
\label{eq:sgata}
\end{equation}
where  the subscript  $J$  labels  the error  space  spanned by  basis
$\{F_i|J_L\rangle\}$ of the  $J$th code word ($F_i$  being the allowed
errors), with suitable pairing $\{m,n\}$,  i.e., one that ensures that
$S_{mm} = S_{nn}^\ast$.  The term  within the curly braces defines the
Hamiltonian $H_S$ suitable to generate $S^+$.

Any   correctable   pure   state    is   an   eigenstate   of   $S^+$:
$S^+|\Psi_L^m\rangle =  S^+\left( \sum_J \alpha_j |J^m_L\rangle\right)
=   \sum_J  \alpha_J   e^{i\theta_m}  |J_L^m\rangle   =  e^{i\theta_m}
|\Psi_L^m\rangle$.      We     thus     have     $S^+\left[\mathcal{E}
  (\rho)\right](S^+)^\dag = \sum_{m,n}\chi_{m,n} S^+|\Psi_L^{m}\rangle
\langle       \Psi^{n}_L|(S^+)^\dag       =       \sum_{m,n}\chi_{m,n}
e^{i(\theta_m-\theta_n)}   |\Psi_L^m\rangle   \langle\Psi^n_L|  \equiv
\sum_{m,n}   \chi^\prime_{m,n}   |\Psi_L^m\rangle   \langle\Psi^n_L|$.
Thus, under  the action of $S^+$, $\chi  \longrightarrow \chi^\prime =
S\chi S^\dag$,  which leaves the  diagonal terms of  $\chi$ invariant,
while the  real and  imaginary parts of  the off-diagonal  elements of
term  $\chi_{m,n}^\prime$ are  interchanged if  $\theta_m =-\theta_n$,
but  are  invariant  otherwise  ($\theta_m=\theta_n$).   Therefore,  a
syndrome measurement  following an application of suitable  $U$ on the
`toggled'  (i.e., $S^+$-applied) noisy  state can  reveal the  real or
imaginary part of $\chi_{j,k}$ that is inaccessible otherwise. 

For a  given $U$,  we determine $d^2$  off-diagonal real  or imaginary
terms  without  $S^+$.   Now  there  exists  a  $S^+$  such  that  the
configuration   $US^+$   suffices   to   cover   all   the   remaining
real/imaginary counterparts  of these terms. This  follows from noting
that these  $d^2$ terms  can be represented  graph theoretically  by a
cyclic  graph  with  $d^2$  vertices,  where  correctable  errors  are
vertices, and edges are pairs of errors that occur in the off-diagonal
terms.  The required  $S^+$ exists precisely because an  even cycle is
always   two-vertex  colorable.    For   example,   suppose  the   $U$
configuration determines  the real or imaginary  parts of $\chi_{1,2},
\chi_{2,3},\cdots,  \chi_{d^2,1}$,  then  in  Eq.   (\ref{eq:S+}),  we
choose $\theta_1 = \theta_3  = \cdots = \theta_{d^2-1}= \frac{\pi}{4}$
and $\theta_2 = \theta_4 = \cdots = \theta_{d^2} = \frac{-\pi}{4}$.

Now one configuration  is enough to determine  all $d^2-1$ independent
diagonal  terms.  This  leaves  $d^4-2d^2+1  = (d^2-1)^2$  independent
off-diagonal  terms  to  be  determined,   for  which  the  number  of
configurations,   $N_{\rm  exp}$   is   at  most   $2  \times   \lceil
\frac{(d^2-1)^2}{d^2}\rceil   =   2(d^2-1)=2(4^n-1)$,  including   the
experiments with  both $U$ and  $US^+$.  This compares  favorably with
SQPT  ($N_{\rm  exp}  =  16^n$), AAPT  with  mutually  unbiased  basis
measurements  ($N_{\rm  exp}=4^n+1$),  and  DSQD  ($N_{\rm  exp}=4^n$)
\cite{ML06}.   As  when  $U$  alone is  applied,  similarly  to  $S^+$
toggling, the correctability is unaffected, allowing the encoded state
to be recovered.  The observed syndrome  will indicate the error to be
corrected, while no information about the encoded state is revealed.

\section{Practical implementation \label{sec:exp}}  

From  an experimental  perspective,  quantum  circuits that  implement
computation can  readily be adapted  into those that  implement QECCD.
For example,  the five-qubit  QECCD code  differs from  the five-qubit
code  of  Ref.   \cite{LMP+96}  only  in  the   choice  of  stabilizer
measurements, and  not in  the encoding.  For an  $[[n,k]]$-qubit code
that performs QECCD  on an $m$-qubit noise, the  quantum Hamming bound
may be stated as
\begin{equation}
2^k|\mathbb{E}| = 2^k 4^m \le 2^n,
\label{eq:QHB}
\end{equation}
from which it follows that the  smallest non-trivial code for QECCD is
not  a five-qubit  code, but  a three-qubit  code, setting  $k=m=1$ in
inequality  (\ref{eq:QHB}).  Thus  a  suitable starting  place for  an
experimental  implementation  of  our   idea  is  a  three-qubit  code
(discussed in  detail below), or  an adaption of the  five-qubit code.
One can  devise a family  of codes that satisfy  bound (\ref{eq:QHB}),
and  correspondingly  a  family  of new  experiments.   QECCD  can  be
implemented with technologies like  NMR \cite{SRM0} a nd linear-optics
with post-selection \cite{klm} that are used for quantum computation.

\begin{figure}
\includegraphics[width=0.4\textwidth]{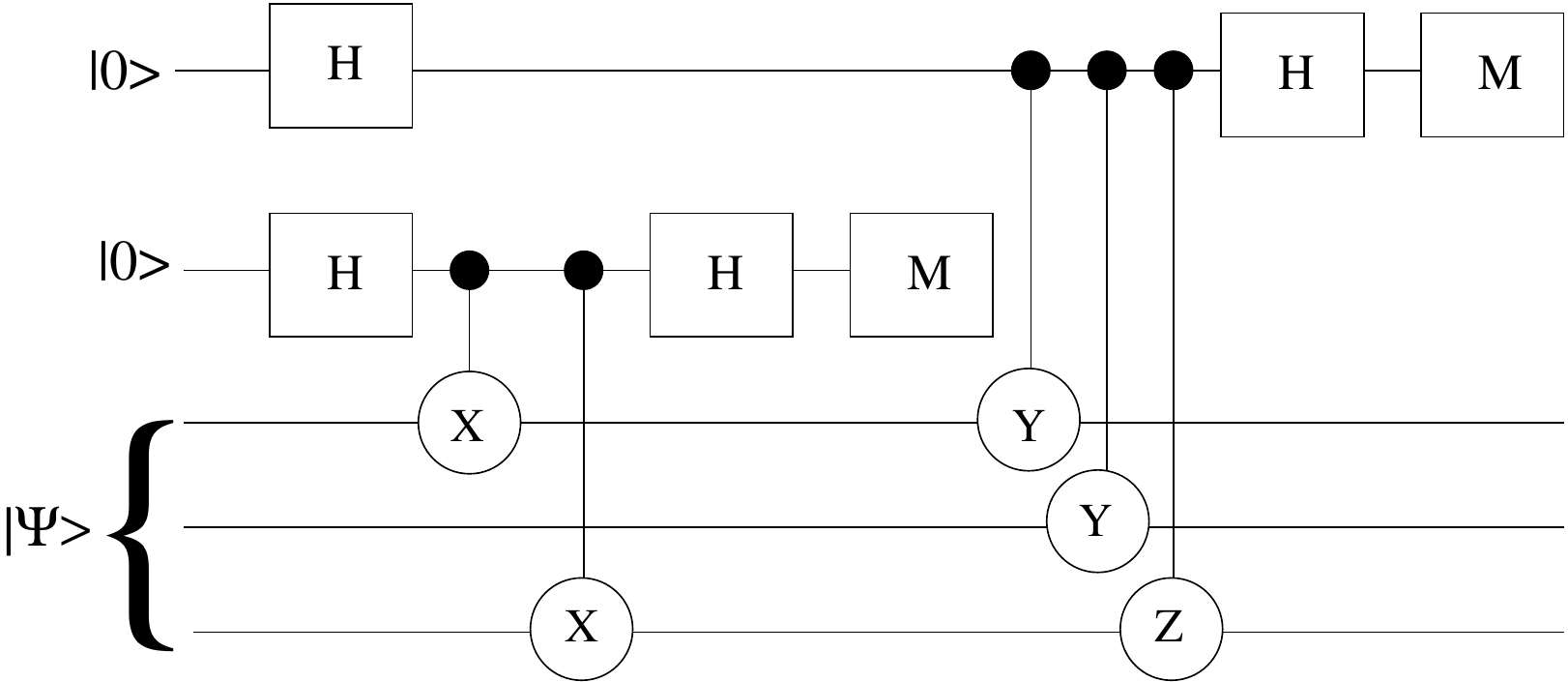}
\caption{Circuit to measure the  stabilizer generators $XIX$ and $YYZ$
  for the $[[3,1]]$ QECC, (\ref{eq:3qec}).   The top two wires are the
  error correction ancillas, while the bottom three wires are the code
  qubits.   Time flows  from  left to  right. The  boxes  $H$ and  $M$
  represent  a Hadamard  and measurement  in the  computational basis,
  respectively.   The circle  with an  operation $U  \in \{X,  Y, Z\}$
  represents  a  control-$U$ operation,  with  control  at the  filled
  circle  on  the   other  end  of  the   ``stick''.   Only  two-qubit
  interactions are used.}
\label{fig:xix}
\end{figure}

Accordingly, let us consider  a one-qubit system \textbf{P}, subjected
to an arbitrary CP channel.  The  Hamming bound is reached with $n=3$,
and a  $[[3,1]]$ QECC (with  qubits 2  and 3 constituting  CQD ancilla
\textbf{A}) that meets the requirement is:
\begin{eqnarray}
|0_L\rangle &=& \frac{1}{2}(|001\rangle  + |010\rangle + |100\rangle +
|111\rangle)  \nonumber \\  |1_L\rangle &=&  \frac{1}{2}(|110\rangle -
|101\rangle + |011\rangle - |000\rangle),
\label{eq:3qec}
\end{eqnarray}
whose stabilizer generators are $XIX$  and $YYZ$, which constitute the
set $\mathcal{G}_3$.  The logical operators  are $X_L \equiv -ZXZ$ and
$Z_L  \equiv XYX$.   We  consider applying  QECCD  to characterize  an
amplitude  damping channel,  determined by  two Kraus  operators, $E_0
\equiv                \frac{1+\sqrt{1-\lambda}}{2}I_2                +
\frac{1-\sqrt{1-\lambda}}{2}Z$         and         $E_1         \equiv
\frac{\sqrt{\lambda}}{2}X    +   \frac{i\sqrt{\lambda}}{2}Y$,    where
$\lambda$, the unknown parameter, is  a measure of the vacuum coupling
strength.  Figure  \ref{fig:xix} depicts the implementation  of one of
the stabilizers for the code.

The  state  $\rho_i  \equiv  |\Psi_L\rangle\langle\Psi_L|$  transforms
under this channel, as per Eq.  (\ref{eq:chi}), to $\rho_f =\sum_{m,n}
\chi_{m,n}|\Psi^m_L\rangle\langle\Psi^n_L|        =        \frac{1}{4}
    [(2-\lambda+2\sqrt{1-\lambda})    |\Psi_L\rangle\langle\Psi_L|   +
      (2-\lambda-2\sqrt{1-\lambda}) |\Psi^Z_L\rangle\langle\Psi^Z_L| +
      \lambda(|\Psi_L\rangle\langle\Psi^Z_L|                          +
      |\Psi^Z_L\rangle\langle\Psi_L|                                  +
      |\Psi^X_L\rangle\langle\Psi^X_L|                                +
      |\Psi^Y_L\rangle\langle\Psi^Y_L|
      -i|\Psi^X_L\rangle\langle\Psi^Y_L|                              +
      i|\Psi^Y_L\rangle\langle\Psi^X_L|)].$  Syndrome  measurements on
    this  state  yield  the   diagonal  terms  of  $\chi$  as  outcome
    probabilities.   The  only  nonvanishing  off-diagonal  terms  are
    $\chi_{I,Z} =  \chi_{Z,I} = \lambda$  and $\chi_{X,Y}=-\chi_{X,Y}=
    -i\lambda$.

Suppose  $U =  U_{X,Y} \equiv  \frac{X +  Y}{\sqrt{2}}$ is  applied to
$\rho_f$, followed by measurement  of the above two stabilizers.  From
Eq.     (\ref{eq:offdiag}),   we   see    that   this    will   reveal
$\textrm{Re}(\chi_{X,Y})=0$ in  the case of  outcomes corresponding to
errors  $I$ and $Z$,  and $\textrm{Im}(\chi_{I,Z})=0$  in the  case of
outcomes  corresponding  to errors  $X$  and  $Y$,  so that  $\lambda$
remains     undetermined.      To     obtain     information     about
$\textrm{Re}(\chi_{X,Y})$  or  $\textrm{Im}(\chi_{I,Z})$, one  applies
prior to  $U(X,Y)$, a toggling operation, which  in the representation
of the basis  $\{|0_L\rangle, |0^x_L\rangle, |0^y\rangle, |0^z\rangle,
|1_L\rangle, |1^x_L\rangle,  |1^y\rangle, |1^z\rangle\}$, is  given by
the diagonal $8 \times 8$ matrix:
\begin{equation}
S^+  \equiv
\frac{1}{\sqrt{2}}\left( \begin{array}{c|c} S & 0 
\\ \hline
0 & S 
\end{array}\right),
\end{equation}
where   $S   =    \textrm{Diag}(\gamma,   \overline{\gamma},   \gamma,
\overline{\gamma})$,  with $\gamma  =  1+i$  and $\overline{\gamma}  =
1-i$.   For the  toggled  channel, $\chi^\prime_{I,J}  = \left(S  \chi
S\right)_{I,J}    =     i\chi_{I,J}    =     i\lambda$.     Similarly,
$\chi^\prime_{X,Y} = i\chi_{X,Y}  = \lambda$.  Thus the  full noise is
determined.  The following three configurations  are used for CQD: (i)
immediate  stabilizer measurement;  (ii) pre-processing  with $U(X,Y)$
before  stabilizer measurement;  (iii) pre-processing  with $S^+$  and
then $U(X,Y)$ before stabilizer measurement.

\section{Discussion and Conclusions \label{sec:conclu}}  

We have proposed QECCD, a method for CQD that exploits QEC techniques.
Like DCQD  \cite{ML06}, CQD is direct  and requires a separation  of a
clean qubit system \textbf{A} from  the noisy qubit system \textbf{P}.
While this  assumption is a bit  restrictive, it is worth  noting that
all AAPT methods (to our knowledge) require this assumption. Expanding
our method so that some noise is allowed in the ancillary system would
be an  interesting future  direction of  work.  Another  direction for
expansion of  our method would  be to incorporate  fault-tolerance, by
allowing the  gate operations  performed during  CQD to  be imperfect.
This approach  may either aim  to determine  a threshold for  the gate
fidelity that would allow CQD to  be accurate, or relate gate fidelity
to the variance in the estimated noise parameters.

Unlike earlier CQD techniques, the QECCD protocol is not restricted to
a  fixed set  of  initial states,  but accepts  as  input any  encoded
quantum information  and, thus,  can be implemented  concurrently with
the QCC.   This has the economizing  virtue that a quantum  state used
for CQD need not be  discarded from the quantum computation procedure.
Moreover, QECCD requires at most only twice the number of experimental
configurations  as does  DCQD  or AAPT  with  mutually unbiased  basis
measurements.   Unlike  AAPT  with  POVMs,  which  requires  many-body
interactions,  QECCD,  like  DCQD,  requires only  one-  and  two-body
interactions \cite{12body}.

We now  highlight some  other insights that  our method  has provided:
First,  we   present  in   Eq.   (\ref{eq:CJ})  a   new  channel-state
isomorphism, which  is similar  to the  Choi-Jamiolkowski isomorphism,
but  with  an  interesting   twist.   To  turn  the  Choi-Jamiolkowski
isomorphism  into  a  method  for  CQD,  one  requires  a  full  state
tomography  on the  state  obtained under  the  isomorphism (the  Choi
matrix).  By our  method, however, the QECC  isomorphism only requires
partial information (syndrome outcome data) at each step.

Second, QEC  involves destroying  the coherence between  Pauli errors,
and only  gives the  probability of  those errors.   Nevertheless, the
QECC isomorphism, (\ref{eq:CJ}), implies  that the noisy encoded state
contains these  coherence data, and  raises the question  whether this
coherence information can be physically accessed using QEC techniques.
Our method answers this question in the affirmative.

Third, some of the mathematical tools  we propose, such as toggling to
determine the  real and imaginary  parts of the off-diagonal  terms of
the density operator, can be of independent interest.

Last  but not  least,  the QECCs  which we  introduced  for QECCD  are
different  in that  they are  novel.  They  are stabilizer  codes that
correct arbitrary errors  on known coordinates, and  have the property
that  the  set   of  allowed  Pauli  errors  forms   a  group.   Codes
(\ref{eq:5qecc})  and  (\ref{eq:3qec})  are  examples  of  such  QECCs
suitable for QECCD.  Interestingly, both these codes are perfect.

\bibliography{QECPT}

%merlin.mbs apsrev4-1.bst 2010-07-25 4.21a (PWD, AO, DPC) hacked
%Control: key (0)
%Control: author (0) dotless jnrlst
%Control: editor formatted (1) identically to author
%Control: production of article title (0) allowed
%Control: page (1) range
%Control: year (0) verbatim
%Control: production of eprint (0) enabled
\begin{thebibliography}{19}%
\makeatletter
\providecommand \@ifxundefined [1]{%
 \@ifx{#1\undefined}
}%
\providecommand \@ifnum [1]{%
 \ifnum #1\expandafter \@firstoftwo
 \else \expandafter \@secondoftwo
 \fi
}%
\providecommand \@ifx [1]{%
 \ifx #1\expandafter \@firstoftwo
 \else \expandafter \@secondoftwo
 \fi
}%
\providecommand \natexlab [1]{#1}%
\providecommand \enquote  [1]{``#1''}%
\providecommand \bibnamefont  [1]{#1}%
\providecommand \bibfnamefont [1]{#1}%
\providecommand \citenamefont [1]{#1}%
\providecommand \href@noop [0]{\@secondoftwo}%
\providecommand \href [0]{\begingroup \@sanitize@url \@href}%
\providecommand \@href[1]{\@@startlink{#1}\@@href}%
\providecommand \@@href[1]{\endgroup#1\@@endlink}%
\providecommand \@sanitize@url [0]{\catcode `\\12\catcode `\$12\catcode
  `\&12\catcode `\#12\catcode `\^12\catcode `\_12\catcode `\%12\relax}%
\providecommand \@@startlink[1]{}%
\providecommand \@@endlink[0]{}%
\providecommand \url  [0]{\begingroup\@sanitize@url \@url }%
\providecommand \@url [1]{\endgroup\@href {#1}{\urlprefix }}%
\providecommand \urlprefix  [0]{URL }%
\providecommand \Eprint [0]{\href }%
\providecommand \doibase [0]{http://dx.doi.org/}%
\providecommand \selectlanguage [0]{\@gobble}%
\providecommand \bibinfo  [0]{\@secondoftwo}%
\providecommand \bibfield  [0]{\@secondoftwo}%
\providecommand \translation [1]{[#1]}%
\providecommand \BibitemOpen [0]{}%
\providecommand \bibitemStop [0]{}%
\providecommand \bibitemNoStop [0]{.\EOS\space}%
\providecommand \EOS [0]{\spacefactor3000\relax}%
\providecommand \BibitemShut  [1]{\csname bibitem#1\endcsname}%
\let\auto@bib@innerbib\@empty
%</preamble>
\bibitem [{\citenamefont {Nielsen}\ and\ \citenamefont {Chuang}(2000)}]{NC00}%
  \BibitemOpen
  \bibfield  {author} {\bibinfo {author} {\bibfnamefont {M.}~\bibnamefont
  {Nielsen}}\ and\ \bibinfo {author} {\bibfnamefont {I.}~\bibnamefont
  {Chuang}},\ }\href@noop {} {\emph {\bibinfo {title} {Quantum Computation and
  Quantum Information}}}\ (\bibinfo  {publisher} {Cambridge University Press
  (Cambridge)},\ \bibinfo {year} {2000})\BibitemShut {NoStop}%
\bibitem [{\citenamefont {D'Ariano}(2000)}]{DAr00}%
  \BibitemOpen
  \bibfield  {author} {\bibinfo {author} {\bibfnamefont {Giacomo~Mauro}\
  \bibnamefont {D'Ariano}},\ }\bibfield  {title} {\enquote {\bibinfo {title}
  {Quantum tomography: General theory and new experiments},}\ }\href {\doibase
  10.1002/(sici)1521-3978(200005)48:5/7<579::aid-prop579>3.0.co;2-p} {\bibfield
   {journal} {\bibinfo  {journal} {F. der Physik}\ }\textbf {\bibinfo {volume}
  {48}},\ \bibinfo {pages} {579--588} (\bibinfo {year} {2000})}\BibitemShut
  {NoStop}%
\bibitem [{\citenamefont {Altepeter}\ \emph {et~al.}(2003)\citenamefont
  {Altepeter}, \citenamefont {Branning}, \citenamefont {Jeffrey}, \citenamefont
  {Wei}, \citenamefont {Kwiat}, \citenamefont {Thew}, \citenamefont {O'Brien},
  \citenamefont {A.Nielsen},\ and\ \citenamefont {White}}]{ABJ+03}%
  \BibitemOpen
  \bibfield  {author} {\bibinfo {author} {\bibfnamefont {J.~B.}\ \bibnamefont
  {Altepeter}}, \bibinfo {author} {\bibfnamefont {D.}~\bibnamefont {Branning}},
  \bibinfo {author} {\bibfnamefont {E.}~\bibnamefont {Jeffrey}}, \bibinfo
  {author} {\bibfnamefont {T.~C.}\ \bibnamefont {Wei}}, \bibinfo {author}
  {\bibfnamefont {P.~G.}\ \bibnamefont {Kwiat}}, \bibinfo {author}
  {\bibfnamefont {R.~T.}\ \bibnamefont {Thew}}, \bibinfo {author}
  {\bibfnamefont {J.~L.}\ \bibnamefont {O'Brien}}, \bibinfo {author}
  {\bibfnamefont {M.}~\bibnamefont {A.Nielsen}}, \ and\ \bibinfo {author}
  {\bibfnamefont {A.~G.}\ \bibnamefont {White}},\ }\bibfield  {title} {\enquote
  {\bibinfo {title} {Ancilla-assisted quantum process tomography},}\ }\href
  {\doibase 10.1103/PhysRevLett.90.193601} {\bibfield  {journal} {\bibinfo
  {journal} {Phys. Rev. Lett.}\ }\textbf {\bibinfo {volume} {90}},\ \bibinfo
  {pages} {193601} (\bibinfo {year} {2003})}\BibitemShut {NoStop}%
\bibitem [{\citenamefont {Mohseni}\ and\ \citenamefont {Lidar}(2006)}]{ML06}%
  \BibitemOpen
  \bibfield  {author} {\bibinfo {author} {\bibfnamefont {M.}~\bibnamefont
  {Mohseni}}\ and\ \bibinfo {author} {\bibfnamefont {D.~A.}\ \bibnamefont
  {Lidar}},\ }\bibfield  {title} {\enquote {\bibinfo {title} {Direct
  characterization of quantum dynamics},}\ }\href {\doibase
  10.1103/PhysRevLett.97.170501} {\bibfield  {journal} {\bibinfo  {journal}
  {Phys. Rev. Lett.}\ }\textbf {\bibinfo {volume} {97}},\ \bibinfo {pages}
  {170501} (\bibinfo {year} {2006})}\BibitemShut {NoStop}%
\bibitem [{\citenamefont {Mohseni}\ and\ \citenamefont {Lidar}(2007)}]{ML07}%
  \BibitemOpen
  \bibfield  {author} {\bibinfo {author} {\bibfnamefont {M.}~\bibnamefont
  {Mohseni}}\ and\ \bibinfo {author} {\bibfnamefont {D.~A.}\ \bibnamefont
  {Lidar}},\ }\bibfield  {title} {\enquote {\bibinfo {title} {Direct
  characterization of quantum dynamics: General theory},}\ }\href {\doibase
  10.1103/PhysRevA.75.062331} {\bibfield  {journal} {\bibinfo  {journal} {Phys.
  Rev. A}\ }\textbf {\bibinfo {volume} {75}},\ \bibinfo {pages} {062331}
  (\bibinfo {year} {2007})}\BibitemShut {NoStop}%
\bibitem [{\citenamefont {Emerson}\ \emph {et~al.}(2007)\citenamefont {Emerson}
  \emph {et~al.}}]{ESM+07}%
  \BibitemOpen
  \bibfield  {author} {\bibinfo {author} {\bibfnamefont {Joseph}\ \bibnamefont
  {Emerson}} \emph {et~al.},\ }\bibfield  {title} {\enquote {\bibinfo {title}
  {Symmetrized characterization of noisy quantum processes},}\ }\href {\doibase
  10.1126/science.1145699} {\bibfield  {journal} {\bibinfo  {journal}
  {Science}\ }\textbf {\bibinfo {volume} {317}},\ \bibinfo {pages} {1893--1896}
  (\bibinfo {year} {2007})}\BibitemShut {NoStop}%
\bibitem [{\citenamefont {Silva}\ \emph {et~al.}(2008)\citenamefont {Silva},
  \citenamefont {Magesan}, \citenamefont {Kribs},\ and\ \citenamefont
  {Emerson}}]{SMK+08}%
  \BibitemOpen
  \bibfield  {author} {\bibinfo {author} {\bibfnamefont {M.}~\bibnamefont
  {Silva}}, \bibinfo {author} {\bibfnamefont {E.}~\bibnamefont {Magesan}},
  \bibinfo {author} {\bibfnamefont {D.~W.}\ \bibnamefont {Kribs}}, \ and\
  \bibinfo {author} {\bibfnamefont {J.}~\bibnamefont {Emerson}},\ }\bibfield
  {title} {\enquote {\bibinfo {title} {Scalable protocol for identification of
  correctable codes},}\ }\href {\doibase 10.1103/PhysRevA.78.012347} {\bibfield
   {journal} {\bibinfo  {journal} {Phys. Rev. A}\ }\textbf {\bibinfo {volume}
  {78}},\ \bibinfo {pages} {012347} (\bibinfo {year} {2008})}\BibitemShut
  {NoStop}%
\bibitem [{\citenamefont {Fowler}\ \emph {et~al.}()\citenamefont {Fowler},
  \citenamefont {Sank}, \citenamefont {Kelly}, \citenamefont {Barends},\ and\
  \citenamefont {Martinis}}]{FSK+14}%
  \BibitemOpen
  \bibfield  {author} {\bibinfo {author} {\bibfnamefont {Austin~G.}\
  \bibnamefont {Fowler}}, \bibinfo {author} {\bibfnamefont {D.}~\bibnamefont
  {Sank}}, \bibinfo {author} {\bibfnamefont {J.}~\bibnamefont {Kelly}},
  \bibinfo {author} {\bibfnamefont {R.}~\bibnamefont {Barends}}, \ and\
  \bibinfo {author} {\bibfnamefont {John~M.}\ \bibnamefont {Martinis}},\
  }\href@noop {} {}\bibinfo {note} {ArXiv:1405.1454}\BibitemShut {NoStop}%
\bibitem [{\citenamefont {Combes}\ \emph {et~al.}()\citenamefont {Combes},
  \citenamefont {Ferrie}, \citenamefont {Cesare}, \citenamefont {Tiersch},
  \citenamefont {Milburn}, \citenamefont {Briegel},\ and\ \citenamefont
  {Caves}}]{CFC+14}%
  \BibitemOpen
  \bibfield  {author} {\bibinfo {author} {\bibfnamefont {J.}~\bibnamefont
  {Combes}}, \bibinfo {author} {\bibfnamefont {C.}~\bibnamefont {Ferrie}},
  \bibinfo {author} {\bibfnamefont {C.}~\bibnamefont {Cesare}}, \bibinfo
  {author} {\bibfnamefont {M.}~\bibnamefont {Tiersch}}, \bibinfo {author}
  {\bibfnamefont {G.~J.}\ \bibnamefont {Milburn}}, \bibinfo {author}
  {\bibfnamefont {H.~J.}\ \bibnamefont {Briegel}}, \ and\ \bibinfo {author}
  {\bibfnamefont {C.~M.}\ \bibnamefont {Caves}},\ }\href@noop {} {}\bibinfo
  {note} {ArXiv:1405.5656}\BibitemShut {NoStop}%
\bibitem [{\citenamefont {Fujiwara}()}]{Fuj14}%
  \BibitemOpen
  \bibfield  {author} {\bibinfo {author} {\bibfnamefont {Y.}~\bibnamefont
  {Fujiwara}},\ }\href@noop {} {}\bibinfo {note} {ArXiv:1405.6267}\BibitemShut
  {NoStop}%
\bibitem [{\citenamefont {Bendersky}\ \emph {et~al.}(2008)\citenamefont
  {Bendersky}, \citenamefont {Pastawski},\ and\ \citenamefont {Paz}}]{BPP08}%
  \BibitemOpen
  \bibfield  {author} {\bibinfo {author} {\bibfnamefont {Ariel}\ \bibnamefont
  {Bendersky}}, \bibinfo {author} {\bibfnamefont {Fernando}\ \bibnamefont
  {Pastawski}}, \ and\ \bibinfo {author} {\bibfnamefont {Juan~Pablo}\
  \bibnamefont {Paz}},\ }\bibfield  {title} {\enquote {\bibinfo {title}
  {Selective and efficient estimation of parameters for quantum process
  tomography},}\ }\href {\doibase 10.1103/PhysRevLett.100.190403} {\bibfield
  {journal} {\bibinfo  {journal} {Phys. Rev. Lett.}\ }\textbf {\bibinfo
  {volume} {100}},\ \bibinfo {pages} {190403} (\bibinfo {year}
  {2008})}\BibitemShut {NoStop}%
\bibitem [{\citenamefont {Shabani}\ and\ \citenamefont {Lidar}(2009)}]{SL09}%
  \BibitemOpen
  \bibfield  {author} {\bibinfo {author} {\bibfnamefont {Alireza}\ \bibnamefont
  {Shabani}}\ and\ \bibinfo {author} {\bibfnamefont {Daniel~A.}\ \bibnamefont
  {Lidar}},\ }\bibfield  {title} {\enquote {\bibinfo {title} {Maps for general
  open quantum systems and a theory of linear quantum error correction},}\
  }\href {\doibase 10.1103/PhysRevA.80.012309} {\bibfield  {journal} {\bibinfo
  {journal} {Phys. Rev. A}\ }\textbf {\bibinfo {volume} {80}},\ \bibinfo
  {pages} {012309} (\bibinfo {year} {2009})}\BibitemShut {NoStop}%
\bibitem [{\citenamefont {Gottesman}()}]{Got09}%
  \BibitemOpen
  \bibfield  {author} {\bibinfo {author} {\bibfnamefont {D.}~\bibnamefont
  {Gottesman}},\ }\href@noop {} {}\bibinfo {note} {ArXiv:0904.2557}\BibitemShut
  {NoStop}%
\bibitem [{\citenamefont {Omkar}\ \emph {et~al.}(2013)\citenamefont {Omkar},
  \citenamefont {Srikanth},\ and\ \citenamefont {Banerjee}}]{OSB0}%
  \BibitemOpen
  \bibfield  {author} {\bibinfo {author} {\bibfnamefont {S.}~\bibnamefont
  {Omkar}}, \bibinfo {author} {\bibfnamefont {R.}~\bibnamefont {Srikanth}}, \
  and\ \bibinfo {author} {\bibfnamefont {Subhashish}\ \bibnamefont
  {Banerjee}},\ }\bibfield  {title} {\enquote {\bibinfo {title} {Dissipative
  and non-dissipative single-qubit channels: dynamics and geometry},}\ }\href
  {\doibase 10.1007/s11128-013-0628-3} {\bibfield  {journal} {\bibinfo
  {journal} {Quant. Info. Proc.}\ }\textbf {\bibinfo {volume} {12}},\ \bibinfo
  {pages} {3725} (\bibinfo {year} {2013})}\BibitemShut {NoStop}%
\bibitem [{\citenamefont {Laflamme}\ \emph {et~al.}(1996)\citenamefont
  {Laflamme}, \citenamefont {Miquel}, \citenamefont {Paz},\ and\ \citenamefont
  {Zurek}}]{LMP+96}%
  \BibitemOpen
  \bibfield  {author} {\bibinfo {author} {\bibfnamefont {Raymond}\ \bibnamefont
  {Laflamme}}, \bibinfo {author} {\bibfnamefont {Cesar}\ \bibnamefont
  {Miquel}}, \bibinfo {author} {\bibfnamefont {Juan~Pablo}\ \bibnamefont
  {Paz}}, \ and\ \bibinfo {author} {\bibfnamefont {Wojciech~Hubert}\
  \bibnamefont {Zurek}},\ }\bibfield  {title} {\enquote {\bibinfo {title}
  {Perfect quantum error correcting code},}\ }\href {\doibase
  10.1103/PhysRevLett.77.198} {\bibfield  {journal} {\bibinfo  {journal} {Phys.
  Rev. Lett.}\ }\textbf {\bibinfo {volume} {77}},\ \bibinfo {pages} {198--201}
  (\bibinfo {year} {1996})}\BibitemShut {NoStop}%
\bibitem [{\citenamefont {Shukla}\ \emph {et~al.}(2013)\citenamefont {Shukla},
  \citenamefont {Rao},\ and\ \citenamefont {Mahesh}}]{SRM0}%
  \BibitemOpen
  \bibfield  {author} {\bibinfo {author} {\bibfnamefont {Abhishek}\
  \bibnamefont {Shukla}}, \bibinfo {author} {\bibfnamefont {K.~Rama~Koteswara}\
  \bibnamefont {Rao}}, \ and\ \bibinfo {author} {\bibfnamefont {T.~S.}\
  \bibnamefont {Mahesh}},\ }\bibfield  {title} {\enquote {\bibinfo {title}
  {Ancilla-assisted quantum state tomography in multiqubit registers},}\ }\href
  {\doibase 10.1103/PhysRevA.87.062317} {\bibfield  {journal} {\bibinfo
  {journal} {Phys. Rev. A}\ }\textbf {\bibinfo {volume} {87}},\ \bibinfo
  {pages} {062317} (\bibinfo {year} {2013})}\BibitemShut {NoStop}%
\bibitem [{\citenamefont {Knill}\ \emph {et~al.}(2001)\citenamefont {Knill},
  \citenamefont {Laflamme},\ and\ \citenamefont {Milburn}}]{klm}%
  \BibitemOpen
  \bibfield  {author} {\bibinfo {author} {\bibfnamefont {E.}~\bibnamefont
  {Knill}}, \bibinfo {author} {\bibfnamefont {R.}~\bibnamefont {Laflamme}}, \
  and\ \bibinfo {author} {\bibfnamefont {G.~J.}\ \bibnamefont {Milburn}},\
  }\href {\doibase 10.1038/35051009} {\bibfield  {journal} {\bibinfo  {journal}
  {Nature}\ }\textbf {\bibinfo {volume} {409}},\ \bibinfo {pages} {46--52}
  (\bibinfo {year} {2001})}\BibitemShut {NoStop}%
\bibitem [{12b()}]{12body}%
  \BibitemOpen
  \href@noop {} {}\bibinfo {note} {All stabilizer measurements and error
  correction operations can be implemented using one- and two-body interactions
  \cite{GPS+07}. To see that such interactions suffice to implement $U$ and
  $S^+$, consider the unitary version of QEC, which requires only one- and
  two-body interactions and can be represented as: $ \forall_{J,a} \left[
  (F_a|J\rangle)|0\rangle_S \rightarrow (F_a|J\rangle)|a\rangle_S \rightarrow
  |J\rangle|a\rangle\right], $ where the first register is the computer and the
  second holds the syndrome register. To implement $U(a,b)$, one prepares the
  second register (an ancilla) in the state $\frac{1}{\sqrt{2}}(|a\rangle +
  |b\rangle)$ and reverses the above operation. To implement $S^+$ one corrects
  the state of the quantum computer, applies the phase gate $|a\rangle
  \longrightarrow e^{i\theta_a}|a\rangle$ on the second register, and then
  ``uncorrects'' the resulting composite system. For the last step, we invoke
  the result that single qubit gates and CNOT are universal for quantum
  computation \cite{NC00}}\BibitemShut {NoStop}%
\bibitem [{\citenamefont {Gupta}\ \emph {et~al.}(2007)\citenamefont {Gupta},
  \citenamefont {Pathak}, \citenamefont {Srikanth},\ and\ \citenamefont
  {Panigrahi}}]{GPS+07}%
  \BibitemOpen
  \bibfield  {author} {\bibinfo {author} {\bibfnamefont {M.}~\bibnamefont
  {Gupta}}, \bibinfo {author} {\bibfnamefont {A.}~\bibnamefont {Pathak}},
  \bibinfo {author} {\bibfnamefont {R.}~\bibnamefont {Srikanth}}, \ and\
  \bibinfo {author} {\bibfnamefont {P.~K.}\ \bibnamefont {Panigrahi}},\
  }\bibfield  {title} {\enquote {\bibinfo {title} {General circuits for
  indirecting and distributing measurement in quantum computation},}\ }\href
  {\doibase 10.1142/S0219749907003092} {\bibfield  {journal} {\bibinfo
  {journal} {Int. Journal of Quantum Information}\ }\textbf {\bibinfo {volume}
  {5}},\ \bibinfo {pages} {627} (\bibinfo {year} {2007})}\BibitemShut {NoStop}%
\end{thebibliography}%

\end{document}